\documentclass[12pt,a4paper]{article}
\usepackage{a4wide}
\usepackage{latexsym}
\usepackage{epsf}
\usepackage{amssymb}
\begin{document}
\def\be{\begin{equation}}
\def\ee{\end{equation}}
\def\bea{\begin{eqnarray}}
\def\eea{\end{eqnarray}}

\def\pd{\partial}
\def\a{\alpha}
\def\b{\beta}
\def\g{\gamma}
\def\d{\delta}
\def\m{\mu}
\def\n{\nu}
\def \h{\mathcal{H}}
\def \hh{\mathcal{G}}
\def\t{\tau}
\def\p{\pi}
\def\th{\theta}
\def\l{\lambda}
\def\O{\Omega}
\def\r{\rho}
\def\s{\sigma}
\def\e{\epsilon}
  \def\scri{\mathcal{J}}
\def\cM{\mathcal{M}}
\def\tcM{\tilde{\mathcal{M}}}
\def\RR{\mathbb{R}}

\hyphenation{re-pa-ra-me-tri-za-tion}
\hyphenation{trans-for-ma-tions}


\begin{flushright}
IFT-UAM/CSIC-04-53\\
hep-th/0410162\\
\end{flushright}

\vspace{1cm}

\begin{center}

{\bf\Large   Reduction of coupling constants in two-dimensional sigma models.}

\vspace{.5cm}

{\bf Enrique \'Alvarez }

\vspace{.3cm}

\vskip 0.4cm  
 
{\it  Instituto de F\'{\i}sica Te\'orica UAM/CSIC, C-XVI,
and  Departamento de F\'{\i}sica Te\'orica, C-XI,\\
  Universidad Aut\'onoma de Madrid 
  E-28049-Madrid, Spain }

\vskip 0.2cm

\vskip 1cm

{\bf Abstract}

Two-dimensional sigma models in curved target spaces are considered 
in which a relationship between the warp factor and the
dilaton is imposed in a renormalization group invariant way.

\end{center}

\begin{quote}

\end{quote}


\newpage

\setcounter{page}{1}
\setcounter{footnote}{1}
\newpage
\section{Introduction}
This paper deals with some aspects of two-dimensional quantum field theories of $d$ scalar
fields $X^{\m}(x^a)$ (where $1\leq \m,\n,\ldots \leq d$ and
 $1\leq a,b,\ldots \leq 2$) of the form:
\be
S=\frac{1}{4\pi l_s^2}\int d^2 x \sum_{a,b= 1,2} g_{\m\n}(X)\pd_a 
X^{\m}\pd _b X^{\n}\gamma^{ab}
\ee
(Here $l_s$ is an arbitrary length scale). The quantities $g_{\m\n}(X)$ can be viewed as
two-dimensional background fields, depending on the imbeddings $X^{\m}(x^a)$,i.e., as 
generalized couplings.   Two-dimensional sigma 
models are interesting for a variety of reasons.
First of all, they obviously enjoy a beatiful geometrical intepretation in terms of embeddings
of a two-dimensional manifold with metric $\g_{ab}$ in a target d-dimensional manifold
with metric $g_{\m\n}$. This fact was successfully exploited by Friedan (\cite{Friedan}) in
order to compute the renormalization group (RG) equations corresponding to the generalized 
couplings. In many cases, they enjoy physical phenomena like a mass gap, 
and have thus been studied
as simplified versions of gauge theories. 
\par
If we supplement the lagrangian density with two other terms, namely
\be
i \e^{ab}b_{\m\n}(X)\pd_a X^{\m}\pd _b X^{\n}+ l_s^2 R \Phi(X)
\ee
corresponding to an antisymmetric two-dimensional background field, $b_{\m\n}(X)$, 
the Kalb-Ramond field, and 
to another scalar two dimensional background field $\Phi(X)$, the dilaton, which couples  
to the two-dimensional scalar curvature,
$R[\g]$, then the sigma model embodies the universal massless fields
of a string. The constant $l_s$ is then the string length, related to the famous $\a^{\prime}$ 
parameter through $\a^{\prime}=l_s^2$.
\par
It is believed that consistency of the quantum string 
demands that this two-dimensional sigma model is Weyl invariant,
which in turn implies the vanishing of the corresponding beta functions of the physical 
string fields,
viewed as generalized two-dimensional couplings.
\par
These beta functions are only known in the sigma model loop expansion, which
physically is an expansion in powers of $(l_s/l)^2$, where $l$ sets the scale of 
curvature in the
target manifold; that is, a low energy, low curvature expansion.
\par

Recent analysis of Perelman \cite{Perelman}, following previous ideas of Hamilton
(cf. \cite{Milnor} for
 a review) of the problem of the classification of three-dimensional geometries draw on
the analysis of precisely this RG equations, which mathematicians call the Ricci flow:
\be
\frac{\pd}{\pd t}g_{\a\b}\equiv \mu^{-1}\frac{\pd}{\pd\mu^{-1}}
g_{\a\b}=-\b_{\a\b}(g_{\rho\sigma})
\ee
\par
A very simple modification of the flow, due also to Hamilton, allows for another proof of the 
uniformization theorem for two dimensional geometries. The modification is
\be
\frac{\pd}{\pd t} g_{\a\b}=-2 R_{\a\b}+\frac{2\pi\chi}{V} g_{\a\b}
\ee
where $V$ is the volume of the two-dimensional manifold
\be
V\equiv \int \sqrt{g}d^2 x
\ee
and $\chi$ the Euler characteristic
\be
\chi\equiv \frac{1}{2\pi}\int \sqrt{g} R d^2 x
\ee
Hamilton \cite{Hamilton} and others proved that the flow exists 
and converges to its fixed points, where 
\be
R=\frac{2\pi\chi}{ V}
\ee
which are precisely the constant curvature metrics.
This modified flow is usually dubbed {\em normalized} for obvious reasons.
\par

Another modification of the flow allows for arbitary target space diffeomorphisms
along it; this amounts to the replacement:
\bea
&&\b_{\m\n}(g_{\a\b})\rightarrow \b_{\m\n}(g_{\a\b})+ \nabla_{\m}\xi_{\n}+\nabla_{\n}\xi_{\m}
\nonumber\\
&&\b (\Phi)\rightarrow\b(\Phi)+\xi^{\m}\nabla_{\m}\Phi
\eea

The fixed points of this flow; that is, the fixed points of the ordinary flow up to
diffeomorphisms, are called {\em Ricci solitons}.
\par

What we want to explore here is the possibility that there are reductions of coupling constants
(studied in the standard model of particle physics 
by Hill (\cite{Hill}) and Zimmermann (\cite{Zimmermann}));
that is, a hypersurface in parameter space which is left invariant by the RG flow. In order
to determine them, it is clearly necessary to study the sigma model away from its conformal
fixed point.
\par
In the absence of $b$-field (just by simplicity), we shall assume a reduction of the form
\be
g_{\m\n}(X) = g_{\m\n}(\Phi(X))
\ee
that is, the metric expressed in terms of the dilaton. This makes physical sense, because the
dilaton is related to the string coupling constant, $g_s$.
\par
Consistency of the RG flow to one loop in the sigma model expansion is achieved when 

\be
R_{\m\n}+ 2 \nabla_{\m}\nabla_{\n}\Phi\bigg|_{g_{\m\n}=g_{\m\n}(\Phi)}=
\frac{d g_{\m\n}}{d \Phi}\left(\Delta-\frac{1}{2}\nabla^2 \Phi + 
g^{\m\n}\nabla_{\m}\Phi \nabla_{\n}\Phi\right)\bigg|_{g_{\m\n}=g_{\m\n}(\Phi)}
\ee

It is tempting to write
\be\label{delta}
\Delta\equiv\frac{d-26}{6 l_s^2}
\ee
as seems to be indicated by the flat space limit, but we would like to keep an open
mind on this, thinking on the curved Liouville interpretation, and we shall treat it as
a free parameter.

This indeed guarantees that the functional relationship is a first integral of the RG equations.

\section{The flow in warped compactifications.}
The essential property of the two-dimensional RG flow is that it is irreversible, at least
when the target space is compact (cf. \cite{Zamo}). This is guaranteed by 
Zamolodchikov's theorem, which actually ensures that there is a monotonically decreasing
 function (the c-function) defined on the flow, and which equals the central charge at the
conformal points. Much less is know for noncompact spaces, although recent progress has been
reported (\cite{Olinyk}) amounting for instance to the absence of breather modes.
\par
In order to get some intuition in some physically interesting examples,  
we shall study the metric
\be
ds^2= d\rho^2 + a(\rho)\d_{ij}dx^i dx^j
\ee 
This metric is the simplest instance of those which admit in adequate circumstances 
a holographic interpretation
(cf. \cite{Polyakov}, and  \cite{Alvarez} for conformal invariant examples). The field $\rho$,
can be thought of as the Liouville mode, and the warping factor $a(\rho)$ as due to the
peculiar measure associated to the Liouville field.

The RG flow is given by: 
\bea\label{flow}
&&\frac{1}{l_s^2}\frac{\pd g_{\r\r}}{\pd log\,\m}=\left(\frac{a^{\prime}}{a}\right)^2 -2 
\frac{a^{''}}{a}+ 2 \Phi^{''}=0
\nonumber\\
&&\frac{\m}{l_s^2}\frac{\pd a}{\pd \m}\d_{ij}=-\frac{1}{2}\left(a^{''}+
\frac{(a^{\prime})^2}{a}\right)\d_{ij} + 
a^{\prime} 
\Phi^{\prime}\d_{ij}\nonumber\\
&&\frac{\m}{l_s^2}\frac{\pd\Phi}{\pd \m}= \Delta-\frac{1}{2}
(\Phi^{''}+\frac{d-1}{2}\frac{a^{\prime}}{a}\Phi^{\prime})+
(\Phi^{\prime})^2
\eea
where $a^{'}\equiv \frac{da}{d\rho}$ and $\Phi^{'}\equiv \frac{d\Phi}{d\rho}$.
It has been kept $g_{\rho\rho}=1$ along the flow, which means more or less imposing
a coordinate gauge condition on it.
\par
The RG equations (\ref{flow}) do not in general keep invariant a given ansatz, and the whole
system of $d(d+1)/2$ gravitational $g_{\m\n}$ variables, plus $d(d-1)/2$ antisymmetric 
$b_{\m\n}$, plus the dilaton $\Phi$, that is, a grand total of $d^2+1$ background
fields, is excited with generic initial conditions.
 \par
In some particular cases, however, it is consistent to assume that some of those fields vanish,
or else keep a simple functional dependence on the imbeddings.
\subsection{Perturbing the anti-de Sitter background}

Let us for example study now in detail (euclidean) anti de-Sitter (adS) geometry,
characterized by:
\be
a(\r)=e^{-\frac{2\r}{l}}
\ee
The first equation of the set (\ref{flow}) then implies that
\be
\Phi= (\frac{\r}{l}+\phi )^2
\ee
(where $\phi$ is a constant)
thus completely determining the dilaton. The two remaining equations 
are consistent only when
\be
\Delta=\frac{1}{l^2}
\ee
and
\be
d=5
\ee
In case $\Delta\equiv
\frac{d-26}{6 l_s^2}$ the radius of adS is given through:
\be
\frac{l^2}{l_s^2}=\frac{6}{d-26}.
\ee
which means that the total dimension must exceed 26.This can be achieved by adding enough 
spectator dimensions.
\par
At any rate, the RG equations then reduce to
\be
\frac{\m}{l_s^2}\frac{\pd\r}{\pd\m}=\frac{2}{l}(1+\phi+\frac{\r}{l})
\ee
which integrates to
\be
\frac{\r}{l}=\bigg(\frac{\m}{\m_0}\bigg)^{\frac{2 l_s^2}{l^2}}-1-\phi
\ee
This leads to the identification of the ultraviolet (UV) region, in the two-dimensional RG
sense, with the region where $\r\rightarrow\infty$, or what is the same, $a\rightarrow 0$.
\subsection{Perturbations of the confining background}
Let us now consider perturbations of the confining background of \cite{Alvarez}. In the 
present notation this is given by:
\bea
&& a(\rho)=\rho+ \d a(\rho)\nonumber\\
&& \Phi(\rho)=\frac{1}{2}log(\r)+\d \Phi(\rho)
\eea
Let us insist again on the fact that it is not guaranteed by any means that there
are perturbations with such a simple dependence on the target space coordinates, and 
with only those background field excited.
In our case, the first equation of (\ref{flow}) imposes  at the linearized level
the restriction
\be
\d a^{\prime\prime}+\frac{\d a}{\r^2}-\frac{\d a^{\prime}}{\r}-\r \d \Phi^{\prime\prime}=0
\ee
Now the other two equations yield two equations for the flow of the two-dimensional
field $\rho$, which for consistency has to be of ${\cal{O}}(1)$, that is, of the same order of
 the perturbation:
\bea
&&\frac{1}{l_s^2}\frac{\pd\r}{\pd\log\,\m}=-\frac{\d a^{\prime\prime}}{2}-
\frac{\d a^{\prime}}{2\r}+ 
\frac{\d a}{2 \r^2}
+\d \Phi^{\prime}\nonumber\\
&&\frac{1}{l_s^2}\frac{\pd\r}{\pd\log\,\m}=-\frac{\d a^{\prime}}{\r}+\frac{\d a}{ \r^2}
-\r\d \Phi^{\prime\prime}
\eea
Consistency of the posited ansatz fully determines the allowed perturbations to be:
\bea
&&\d a = 2 c_1 + c_2 \r + c_3 \r log(\r)\nonumber\\
&&\d \Phi= c_4 + \frac{c_1}{\r}
\eea
Under these circumstances, many terms cancel in the equations, in such a way that the flow
is completely independent of the perturbation of the dilaton, and is only sensitive to the
$\r log(\r)$ piece of the perturbation of the metric. The constant $c_3$ has to be negative,
and the equation can be easily integrated to yield
\be
\rho\sim (log\,(\m/\m_0))^{1/2}
\ee
so that this time we obtain the opposite behavior to the previous example; the UV region 
corresponds to the direction of growing warping factors.

\section{Reduction of coupling constants in warped compactifications.}

Let us now finally examine the possibility that all two-dimensional coupling constants, such as
the spacetime metric, are functions of one amongst them, for example the dilaton.
\par
First of all, in all consistent perturbations with some symmetry, such as the ones considered
in the previous paragraph, this is always so: given the  functions $ a(\r)$ and $\Phi(\r)$
one can always construct the function $a(\Phi)$. What we want to do here is to give a 
direct construction of the general situation. 
\par
\bea\label{redge}
&&\frac{1}{l_s^2}\frac{\pd g_{\r\r}}{\pd \log\,\m}=
\left(\frac{a^{\prime}}{a}\right)^2 -2 \frac{a^{''}}{a}+ 2 \Phi^{''}
\nonumber\\
&&\frac{1}{l_s^2}\frac{\pd a}{\pd \log\,\m}\d_{ij}=-\frac{1}{2}\left(a^{''}+
\frac{(a^{\prime})^2}{a}\right)
\d_{ij} + a^{\prime} 
\Phi^{\prime}\d_{ij}\nonumber\\&&= \frac{d a}{d\Phi}\bigg(\Delta-\frac{1}{2}
(\Phi^{''}+\frac{d-1}{2}\frac{a^{\prime}}{a}\Phi^{\prime})+
(\Phi^{\prime})^2\bigg)\d_{ij}
\eea
Let us consider reduced flows starting from adS. We shall consider $\frac{d a}{d\Phi}=
\frac{a^{'}}{\Phi^{'}}$, although more general situations can be examined, in which
the background fields depend in a nontrivial way on all their arguments.
The second equation of the set (\ref{redge}), which is the nontrivial equation guaranteeing
the fact that the constraint is a first integral of the RG flow, reduces to
\be
\Phi^{\prime\prime}+\frac{5-d}{l}\Phi^{\prime}- 2\Delta=0
\ee
which can be easily solved when $d\neq 5$ to yield:
\be
\Phi= A+\frac{2l\Delta}{5-d}\r+B e^{-\frac{5-d}{l}\r}
\ee
where $A$ and $B$ are arbitrary dimensionless constants.
This is equivalent to the functional relationship
\be
\Phi=-\frac{l^2 \Delta}{5-d}log\, a -B a^{\frac{5-d}{2}} + A
\ee
When $d=5$ instead,
\be
\Phi=\Delta\r^2 + C_1 \frac{\r}{l}+C_2
\ee
where $C_1$ and $C_2$ are dimensionless constants.
\par

The flow is now determined (again, when $d\neq 5$) through
\be
\frac{\pd \r}{\pd log\,\m}=\a+\b e^{-\g \r}
\ee
with $\a\equiv \frac{2 l_s^2}{l}+\frac{2 l \Delta l_s^2}{5-d}$,\quad $\b\equiv \frac{d-5}{l}
B l_s^2$ 
and  $\g\equiv 
\frac{5-d}{l}$

The solution is

\be
e^{-\g\r}=\frac{\a}{D l (\m/\m_0)^{\a\g}-\b}
\ee
where $D$ is a new dimensionless constant of integration.
\par

The character of the flow depends on the signs appearing in this equation. For
example $\a\g=2\frac{l_s^2}{l^2}(5-d+l^2 \Delta)\sim 2 l_s^2 \Delta$ if the curvature
is small in string units. 
\par
There are two types of behavior: either

\be
e^{-\g\r}\sim\frac{1}{(\m/\m_0)^{|C_1|}\pm |C_2|}
\ee
(with $C_1$ and $C_2$ arbitrary constants)
which tends to a constant in the IR, and to zero in the UV, when the
sign in the denominator is positive, or to a Landau pole with the negative sign. 
\par
The other possible behavior is

\be
e^{-\g\r}\sim\frac{1}{(\m/\m_0)^{-|C_1|}\pm |C_2|}
\ee
tending to a constant in the UV and to $0$ in the IR, again with Landau poles appearing when
the sign in the denominator is negative.

\par
Finally, in the special case $d=5$, the flow is
\be
\frac{\r}{l}= E + F \left(\frac{\m}{\m_0}\right)^{2\Delta l^2}
\ee
withe arbitrary dimensionless constants $E$ and $F$. When $\Delta<0$ both the UV and IR 
limits are then
$\r= constant$, which can again be tuned to keep $g_{\r\r}$ bounded.

\subsection{Constrained flow}
Alternatively, it is possible to impose the constraint  that $g_{\rho\rho}=1$ along the flow.
The relevant equations are in this case:
\bea\label{red}
&&\left(\frac{a^{\prime}}{a}\right)^2 -2 \frac{a^{''}}{a}+ 2 \Phi^{''}=0
\nonumber\\
&&-\frac{1}{2}\left(a^{''}+\frac{(a^{\prime})^2}{a}\right)\d_{ij} + a^{\prime} 
\Phi^{\prime}\d_{ij}= \frac{d a}{d\Phi}\bigg(\Delta-\frac{1}{2}
(\Phi^{''}+\frac{d-1}{2}\frac{a^{\prime}}{a}\Phi^{\prime})+
(\Phi^{\prime})^2\bigg)\d_{ij}
\eea
In our simple setting it is still true that
\be
\frac{d a}{d\Phi}=\frac{a^{'}}{\Phi^{'}}
\ee

We shall take the following point of view. Given the dilaton, $\Phi(\rho)$, the warp
factor is determined by the first equation to be:
\be
a=a_0 e^{\int^{\rho}f(x) dx}
\ee
where the auxiliary function $f(x)$ is a solution of the Riccati differential equation:
\be\label{u}
f^{\prime}+\frac{f^2}{2}=\Phi^{\prime\prime}
\ee
We shall write
\be
f[\Phi]
\ee
to emphasize the fact that this function is actually fully determined, once the
dilaton $\Phi$ is known.
\par
The remaining equation is then a very implicit one for the dilaton:
\be
\Phi^{\prime}\left(\frac{f^{\prime}[\Phi]}{2}+\frac{5-d}{4}f[\Phi]^2\right)-\frac{1}{2}f[\Phi]
\Phi^{\prime\prime}
+\Delta f[\Phi]=0
\ee

It is remarkable that a linear dilaton demands $\Delta=0$ and besides, $d=4$ 
\footnote{The apparent contradiction between these two statements when considering 
critical strings can again be resolved by introducing spectator dimensions.}.
The first equation (\ref{u}) then gives $f\sim\frac{2}{\r}$, that is,
\be
a=a_0 (\r/\r_0)^2\sim \Phi^2
\ee

It is clear that a whole taxonomy of reduced flows can be constructed along these lines.
\par
What is more important from the physical point of view is whether there is uniformization
of the two-dimensional RG flow for arbitrary dimension of the target space. This would 
mean that the sigma model
evolves (in the IR) towards the conformal fixed point, and if the reduction of coupling 
constants holds 
all the way during the flow, it should also hold in the conformal limit.

\section{Concluding remarks}
We have already mentioned the remarkable fact 
 that the two dimensional sigma model RG evolution equations allow for a 
different proof 
of the  classification theorem of two-dimensional closed surfaces , and is currently believed
to be the main ingredient towards the proof of Thurston's geometrization conjecture
in three dimensions . Although this problem has been analyzed from a physicist'viewpoint in 
\cite{Gegenberg}\cite{Bakas}, uncovering interesting relationships
with integrable models of the Toda family in the two-dimensional case, 
it is likely that many more fascinating relationships between 
mathematics and physics lie 
hidden beneath the surface.
\par
On the other hand, it is not clear what is the relationship of these models with string theory.
\par
Critical strings corresponds to conformal points, and even non-critical (Liouville) strings
are usually assumed to correspond to conformal models, albeit with non-trivial
background fields. 
\par
Many efforts have been devoted over the years to find a string
 representation of a gauge theory. The main success up to date has been the adS/CFT conjecture
of Maldacena \cite{Maldacena}. Although this particular example corresponds to a non-confining
theory, simple modifications of the original correspondence are able to represent confinement, 
to a certain extent \cite{Witten}. In all these examples, the string dealt with 
is a critical string. 
\par
It is not unthinkable, however, that some relationship  exists
between the four-dimensional 
RG flow of a non supersymmetrical 
gauge theory and the two-dimensional RG flow of some sigma-model related to 
the hypothetical confining string.
\par
The simplest of those models, (such as the ones in \cite{Alvarez}) in which
a direct relationship is posited between the gravitational degrees of freedom and the
gauge coupling constant (determined by the string coupling constant, which is in turn 
determined by the dilaton) are the ones in which a reduction of two-dimensional 
coupling constants is made, somewhat similar to the ones introduced in this paper.
\par
More work is clearly needed in this facinating topic, both by itself, and by 
its possible relationship with the confining string.

\section*{Acknowledgments}
I am indebted to Jorge Conde for a careful reading of the manuscript.
This work has been partially supported by the
European Commission (HPRN-CT-200-00148) and by FPA2003-04597 (DGI del MCyT, Spain).


\end{document}